# An Integrated Tantalum Sulfide – Boron Nitride – Graphene Oscillator: A Charge-Density-Wave Device Operating at Room Temperature

Guanxiong Liu[1], Bishwajit Debnath[2], Timothy R. Pope[3], Tina T. Salguero[3], Roger K. Lake[2] and Alexander A. Balandin[1,*]

[1]Nano-Device Laboratory (NDL) and Phonon Optimized Engineered Materials (POEM) Center, Department of Electrical and Computer Engineering, University of California – Riverside, Riverside, California 92521 USA

[2]Laboratory for Terascale and Terahertz Electronics (LATTE), Department of Electrical and Computer Engineering, University of California – Riverside, Riverside, California 92521 USA

[3]Department of Chemistry, University of Georgia, Athens, Georgia 30602, USA

**The charge-density-wave (CDW) phase is a macroscopic quantum state consisting of a periodic modulation of the electronic charge density accompanied by a periodic distortion of the atomic lattice in quasi-1D or layered 2D metallic crystals[1-4]. Several layered transition metal dichalcogenides, such as $1T$-TaSe$_2$, $1T$-TaS$_2$ and $1T$-TiSe$_2$, exhibit unusually high transition temperatures to different CDW symmetry-reducing phases[1, 5-6]. These transitions can be affected by environmental conditions, film thickness and applied electric bias[1]. However, device applications of these intriguing systems at room temperature or their integration with other 2D materials have not been explored. Here we show that in 2D CDW $1T$-TaS$_2$, the abrupt change in the electrical conductivity and hysteresis at the transition point between nearly-commensurate and incommensurate charge-density-wave phases can be used for**

**constructing an oscillator that operates at room temperature. The hexagonal boron nitride was capped on 1$T$-TaS$_2$ thin film to provide protection from oxidation, and an integrated graphene transistor provides a voltage tunable, matched, low-resistance load enabling precise voltage control of the oscillator frequency. The integration of these three disparate two-dimensional materials, in a way that exploits the unique properties of each, yields a simple, miniaturized, voltage-controlled oscillator device. Theoretical considerations suggest that the upper limit of oscillation frequency to be in the THz regime.**

Main Text

The 1$T$ polytype of TaS$_2$ undergoes the transition from a normal metallic phase to an incommensurate CDW (IC-CDW) phase at 545 K, then to a nearly-commensurate CDW (NC-CDW) phase at 350 K, and, finally, to a commensurate CDW (C-CDW) phase at 180 K[7-11]. Each phase transition is accompanied by a lattice reconstruction, which results in strong modifications of the material's electrical properties. In 1$T$-TaS$_2$, the C-CDW phase drives a Mott-Hubbard metal-to-insulator transition (MIT) that results in a Mott insulator state. The transitions to different CDW phases can be modified by the ambient pressure, thickness of the 2D films, and applied electric bias[12-14]. For example, when the 1$T$-TaS$_2$ film thickness is <10 nm, the C-CDW–NC-CDW transition at 180 K disappears whereas the NC-CDW–IC-CDW transition at 350 K persists[10]. The phase transition also can be triggered by a lateral voltage. In the case of 1$T$-TaS$_2$, its electrical conductivity abruptly increases when the applied electric field exceeds 20 kV/cm, attributed to the insulator-to-metal transition (IMT)[14]. A reverse MIT is revealed when the voltage is swept back[14]. Despite strong recent interest in the CDW properties of 2D transition metal dichalcogenides[10-12, 14-17], neither IMT nor MIT induced by electric field have been observed above room temperature (RT). Correspondingly, no devices that use CDW transitions as the basis of their operation have been demonstrated at or near room temperature.

In this work, we show that an abrupt change in the electrical conductivity and hysteresis observed at the IMT and MIT transitions associated with the transition between the NC-CDW and IC-CDW phases in a 2D 1$T$-TaS$_2$ channel can be used for constructing an oscillator that operates at room temperature and up to 320 K. A second 2D material, hexagonal boron nitride ($h$-BN), provides environmental protection against oxidation and preserves the CDW phase. A third integrated 2D material, graphene, provides a transistor with well-matched current density and load resistance to the TaS$_2$ that allows precise voltage control of the oscillator frequency. The voltage-controlled oscillator demonstrated here does not require a complicated biasing circuit and potentially can lead to inexpensive technologies for communication systems and other high-frequency applications.

High-quality 1$T$-TaS$_2$ crystals were prepared by the chemical vapor transport method (see Methods). The prototype devices were fabricated with thin films mechanically exfoliated from millimeter-sized crystals. To protect the exfoliated 1$T$-TaS$_2$ from oxidation, we used a hexagonal boron nitride ($h$-BN) capping layer positioned on top of the CDW channels, which is a method applicable to diverse 2D materials[18, 19]. The fabricated devices consisted of 1$T$-TaS$_2$ channels with thicknesses in the range of 6–9 nm and Pd/Au (15/60 nm) contacts. An optical image of a typical device is shown in an inset in Fig. 1b (additional data are provided in Supplementary Information S1- S6). The dark purple strip is the 1$T$-TaS$_2$ channel on a SiO$_2$ substrate and the larger green region is the $h$-BN over the 1$T$-TaS$_2$. Fig. 1a shows the current-voltage (I-V) characteristics of such TaS$_2$-BN devices in a temperature range from 78 to 395 K, which includes the 1$T$-TaS$_2$ phase transition from the NC-CDW to IC-CDW states. Below 320 K we observe a prominent abrupt current jump as the voltage exceeds certain value, denoted here as $V_H$. This behavior is also referred to as threshold switching (TS) between the low and high conductive states. When the voltage is swept back to zero, a reverse threshold switch occurs at a smaller voltage $V_L$, which corresponds to a smaller critical electric field. The difference between two voltages forms a hysteresis window. Details extracted from Fig. 1a are shown in Supplementary Information S4. It is important to note that the TS effect in our device exists throughout the temperature range from 78 to 320 K. As the temperature is increased to 345 K, slightly below the NC-CDW–IC-CDW transition at 350 K, TS becomes very weak (see inset to Fig. 1a). Increasing the temperature above 350 K results in the disappearance of TS and the I-V characteristics becomes linear as the 2D 1T-TaS$_2$ channel enters the IC-CDW phase. Fig. 1b presents the temperature-dependent resistance measured with small voltage (20 mV). The resistance decreases as temperature increases until reaching the NC-CDW–IC-CDW transition point near 350 K where the resistance drops abruptly. After the transition the resistance becomes almost constant with temperature. In the cooling cycle, the resistance jump occurs at 340 K, forming a temperature hysteresis. The absence of a C-CDW to NC-CDW phase transition at 180 K in very thin films (<9 nm) of 1$T$-TaS$_2$ is consistent with a recent report[10].

[Figure 1]

Another interesting observation is that the resistance value after electrical-induced switching is the same for all temperatures, as seen from the same slope of I-Vs at high electric field in Fig. 1a. The resistance extracted from the slope is 580 Ω, which matches the low-field temperature dependent resistance 590 Ω presented in Fig. 1b at T> 350 K. In addition, the low-field resistances extracted at lower temperature from the I-V slopes in Fig. 1a are also in agreement with the measured resistance in Fig. 1b, shown in Supplementary Information Fig. S8. This equivalence in resistance values is consistent with the picture of the voltage-controlled switching driving the 1$T$-TaS$_2$ between the NC-CDW the I-CDW phases. Another possibility is that the applied field and the resulting current destroy the Mott insulating state within the individual CDW domains while leaving the lattice in the NC-CDW phase. The microscopic mechanism may have implications for the fundamental switching speed between the low and high resistance states. We also note that the TS voltage scales linearly with the channel length (details in Supplementary Information S5). The critical fields $E_H$ and $E_L$ are on the order of 28 and 23 kV/cm at 78 K and decrease as temperature increases. The length scaling suggests a possibility of further reduction of the threshold voltage for shorter channel devices.

We observe that the I-V characteristic below $V_H$ does not have a simple linear dependence. Rather it has an additional exponential component as shown in Fig. 1c. The temperature dependence of the threshold voltages $V_H$ ($V_L$) follows the relation $(1-T/T_{NC\text{-}IC})^{1/2}$ below the NC-CDW–IC-CDW transition temperature of 350 K (see Fig. 1d). This temperature dependence is the same as described by analytical theories of the CDW order parameter[1]. This temperature dependence has recently been observed in $2H$-TaS$_2$[20].

By connecting this $1T$-TaS$_2$ device with an off-chip $R_L$=1 kΩ load resistor (see schematic in Fig. 2a), we observe stable voltage oscillation at the node between these two components when applying a DC bias $V_{DC}$. These voltage oscillations, monitored by an oscilloscope, are shown in Fig. 2b. The initiation of oscillations requires $V_{DC}$ within a certain range. When $V_{DC}$ is below 3.8 V, no oscillations occur. As $V_{DC}$ increases, the oscillations become sinusoidal. The oscillation frequency, $f$, is dependent on $V_{DC}$. In this specific device, $f$ was tuned from 1.77 to 2 MHz as the $V_{DC}$ changed from 3.83 V to 3.95 V. When $V_{DC}$ becomes greater than 3.98 V, the oscillation stops. The oscillation mechanism can be understood as the switching of the $1T$-TaS$_2$ device between IMT and MIT states with the negative feedback from the load resistor $R_L$. When $V_{DC}$ is large enough, the DC voltage across the $1T$-TaS$_2$ channel reaches $V_H$, and triggers IMT. The sudden increase in current results in a large voltage increase across the load resistor. Since $V_{DC}$ is constant, the instantaneous voltage across the $1T$-TaS$_2$ decreases, driving the 1$T$-TaS$_2$ back into the insulating state. The current then decreases, reducing the voltage across the resistor and increasing the voltage across 1$T$-TaS$_2$, driving it back into its metallic state. As long as this voltage can increase to a level higher than V$_H$ the cycle keeps repeating. Fig. 2c provides additional explanation. The intersection of the 1$T$-TaS$_2$ curve with the load line is the operation DC voltage at the output. The line representing V$_{DC}$=3.8 V is on the edge of V$_H$, where the oscillations just start. Since the oscillation is triggered by IMT–MIT processes, it also can be observed at lower temperature (see Supplementary Information S6).

[Figure 2]

Voltage control of an oscillator frequency using simple circuitry is highly desirable for numerous miniaturized electronic systems. Conventional voltage-controlled oscillators (VCOs) usually require multiple transistors and resistors for biasing[21, 22]. By monolithically integrating a top-gate graphene field-effect transistor (G-FET) with a $h$-BN capped 1$T$-TaS$_2$ device, we demonstrate a VCO operating at RT. The gate and contact electrodes for this 2D integrated TaS$_2$–BN–graphene oscillator were fabricated with electron-beam lithography, dry etch and metal evaporation (Supplementary Information S3). The device structure and scanning electron microscopy image are presented in Fig. 3a and 3b, respectively. The equivalent circuit and the biasing scheme are shown in the inset of Fig. 3a. The DC bias is applied across the integrated structure while the gate voltage of the G-FET is used to change R$_L$ and, hence, tune the oscillation frequency. Note that both the 1$T$-TaS$_2$ and graphene sides of the channel are contacted from the edge, forming quasi-1D contacts (18, 23). The $h$-BN capping entirely covers the 1$T$-TaS$_2$ and graphene sides of the channel to protect against environmental exposure.

[Figure 3]

The resulting device demonstrates excellent control of the oscillations by the gate voltage, $V_G$, of G-FET. The gate voltage can turn oscillations on and off and tune $f$ monotonically. As shown in Fig. 3c, no oscillations occur at the output when $V_G$ is set at zero. As $V_G$ increases to 0.68 V the device starts to produce stable oscillations. For $V_G$ varying from 0.68 V to 1.80 V, the frequency can be tuned monotonically from 1.42 MHz to 1.04 MHz with $V_G$. As $V_G$ increases above 1.80 V, the oscillations begin to show missing cycles, which indicate that the bias setting exceeds the oscillation requirement. $V_{DC}$ is set at 3.65 V in this measurement. The tuning mechanism can be understood by examining the load line. As the slop of the load line changes with $V_G$ and falls within the red hatched region in Fig. 3d the oscillations appear at the output. The transfer characteristic of the G-FET is shown in the inset to Fig. 3d. Note that $V_{DS}$ is biased at 2.4 V, which is close to the value of the voltage drop across G-FET when the circuit is operating at $V_{DC}$=3.65 V. The charge neutrality point in this case is at 2.4 V, shifted by the large $V_{DS}$. As $V_G$ increases from zero to 2.1 V, the resistance of the graphene channel monotonically increases resulting in decreasing oscillation frequency due to the increased RC time constant. Fig. 3e shows the gate voltage tunablity of the oscillator frequency. The tuning sensitivity is 0.3 MHz/V, and the oscillation amplitude $V_{PP}$ is ~0.2 V. Control of both characteristics will improve dramatically as the technology of handling 2D materials advances.

The monolithic integration of a 2D CDW channel with graphene has several particular advantages. The current carrying capacity of G-FETs[24] makes graphene a well-matched load to the low-resistance $TaS_2$. The oscillations occur when IMT is triggered in 1$T$-$TaS_2$ in the mA range (for the 1-2 µm wide samples). With similar dimensions as the 1$T$-$TaS_2$, graphene provides a well-matched current density and resistance. Owing to the linear resistance tunablity with the gate voltage of GFET at high $V_{DS}$ and high carrier concentration, the frequency of the oscillator can be tuned by $V_G$ linearly over a wide range. The tuning linearity of the voltage-controlled oscillator is required for many practical applications[21]. An additional advantage of the CDW–G-FET integrated device is that G-FETs have been demonstrated to operate at very high frequencies of hundreds of GHz[25, 26], which allows for increased oscillation frequency. In our proof-of-concept measurements, the frequency is limited by the extrinsic RC time constants of the probe station measurement apparatus. The intrinsic resistances and capacitances can be reduced by scaling. Graphene and *h*-BN layers in the device structure also act as heat spreaders owing to their high thermal conductivity[27, 28].

From the circuit models in Figs. 2a or 3a, the frequency is determined by the charging and discharging time of the capacitance given by

$$1/f = \mathrm{T} = C\,[(R_s//R_{di})\ln(\alpha_i) - (R_s//R_{dm})\ln(\alpha_m)] \qquad (1)$$

where $\alpha_{m,i} = [V_{DD} - (1+R_s/R_{dm,di})V_L]/[V_{DD} - (1+R_s/R_{dm,di})V_H]$, $R_{di}$ is the $TaS_2$ resistance in the insulating state, $R_{dm}$ is the $TaS_2$ resistance in the metallic state, $R_s$ is the load resistance of either the resistor or G-FET, and *C* is the lumped capacitance between the output node and ground (Supplementary Information S7). The intrinsic geometric capacitances and the resistivites of the 1*T*-$TaS_2$ film and the load resistor or transistor set the upper frequency limit of oscillation. The sheet resistivities of the 1*T*-$TaS_2$ in the high and low resistance states are 3 kΩ/□ and 740 Ω/□, respectively. These values are relatively low allowing for extremely small RC time constants. Considering only the parallel plate capacitance to ground of the $TaS_2$ and graphene, the intrinsic RC limited frequency given by Eq. (1) for the ~ 1 μm scale device shown in Fig. 3b is in the THz regime. Since the RC limited frequency is so high, the intrinsic fundamental limit on the maximum frequency may be set by the time constants determined by the microscopic switching mechanism between the low and high resistance states. The voltage or current driven switching could result from (a) the destruction of the Mott insulating state in the individual NC-CDW domains while leaving the lattice in the NC-CDW state, or it could be (b) accompanied by a reconstruction of the lattice from the NC-CDW state to the I-CDW state. Process (a) is a purely electronic process, and, therefore, extremely fast. The speed of process (b) is governed by the time constant associated with the relaxation of the lattice. Analysis of femtosecond ARPES data at room temperature gives a time constant of 700 fs, which provides an upper frequency limit on the order of 1 THz[29]. In practice, extrinsic resistances and capacitances will determine actual frequency limits.

## Methods

### 1T-TaS2 crystal growth

The source 1*T*-TaS$_2$ crystals were grown by the chemical vapor transport method, where the 1*T* polytype can be isolated by fast quenching from the crystal growth temperature (875-975 °C) [30]. Elemental tantalum (20.4 mmol, Sigma-Aldrich 99.99% purity) and sulfur (41.1 mmol, J.T. Baker >99.9% purity) were ground with mortar/pestle and placed in a 17.8×1.0 cm fused quartz ampule (cleaned with overnight nitric acid soak followed by 24 h anneal at 900 °C). Elemental iodine (J.T. Baker 99.9% purity) was added (~88 mg for a ~14.0 cm$^3$ ampule volume). The ampule was evacuated and backfilled three times with argon, with cooling to mitigate I$_2$ sublimation. Next the ampule was flame sealed and heated in a two-zone tube furnace at 10 °C min$^{-1}$ to 975 °C (hot zone) and 875 °C (cool zone). These temperatures were held for one week. Then the ampule was removed from the hot furnace and immediately quenched in a water-ice-NaCl bath. The resulting golden crystals, shown in Figure S1, were characterized as detailed below. The as-grown crystals and their atomic structure are shown in Supplementary Information Fig. S1. The structure and phase purity were verified by powder X-ray diffraction shown in Supplementary Information Fig. S2, and the stoichiometry was confirmed with energy dispersive spectroscopy and electron-probe microanalysis (shown in Supplementary Information Fig. S3 and Table S1).

### Electrical Characteristics of 1*T*-TaS$_2$ Device

All current-voltage (I-V) characteristics were measured in the Lakeshore cryogenic probe station TTPX with a semiconductor analyzer Agilent B1500. The oscillator of the waveform is collected by oscilloscope LeCroy WaveAce1012 while the DC voltage supply of the circuit is biased with BK Precision 1787B.

**Acknowledgements**

Nanofabrication and device testing were supported, in part, by the National Science Foundation (NSF) and Semiconductor Research Corporation (SRC) Nanoelectronic Research Initiative (NRI) for the project 2204.001: Charge-Density-Wave Computational Fabric: New State Variables and Alternative Material Implementation (NSF ECCS-1124733) as a part of the Nanoelectronics for 2020 and beyond (NEB-2020) program and by the Semiconductor Research Corporation (SRC) and Defense Advanced Research Project Agency (DARPA) through STARnet Center for Function Accelerated nanoMaterial Engineering (FAME). Material synthesis and device simulations were supported by the Emerging Frontiers of Research Initiative (EFRI) 2-DARE project: Novel Switching Phenomena in Atomic $MX_2$ Heterostructures for Multifunctional Applications (NSF 005400).

**Contributions**

A.A.B. coordinated the project and contributed to experimental data analysis; R.K.L. led the theoretical analysis; T.T.S. supervised material synthesis and contributed to materials characterization; G.L. designed, fabricated, tested devices and analyses the experimental data; T.R.P. synthesized $TaS_2$ crystals; B.D. conducted computer simulations. All authors contributed to writing of the manuscript.

G. Liu, B. Debnath, T.R. Pope, T.T. Salguero, R.K. Lake and A. A. Balandin – 2015/2016

**Figure 1**

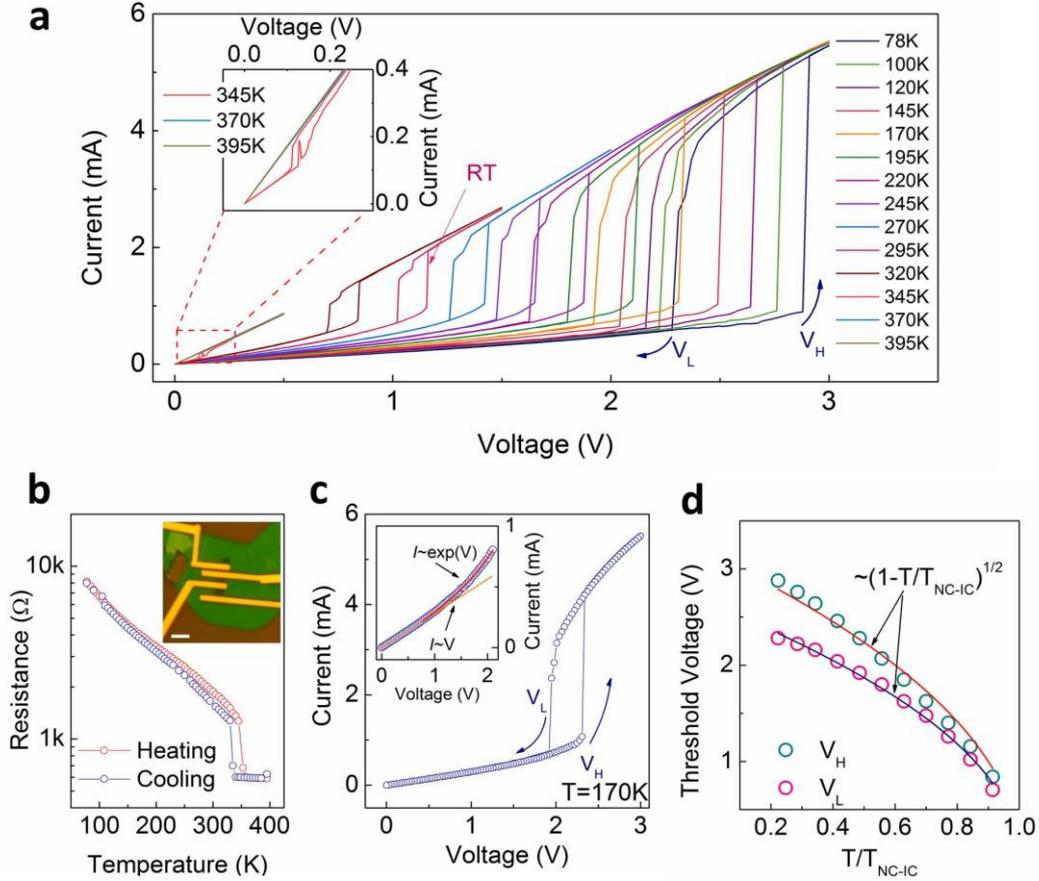

**Figure 1: Electrical characteristics of thin film 1$T$-TaS$_2$. a**, Current-voltage (I-V) characteristics of thin film 1$T$-TaS$_2$ at different temperatures from 78 K to 395 K. The TS effect is prominent from 78 K to 320 K. The blues arrows indicate the voltage sweep direction for the measurement at 78 K. For all the other temperatures, $V_H$ is always higher than $V_L$. The hysteresis window is defined as $V_H$-$V_L$. The TS is prominent up to 320 K, and becomes less pronounced as the temperature approaches the NC-CDW−IC-CDW transition at 350 K. As shown in the inset, at 345 K (red curve), the TS is still measurable. As the temperature exceeds 350 K, the IV becomes linear. **b**, Temperature-dependent resistance measurements for a 9 nm thick 1$T$-TaS$_2$. The NC-CDW−IC-CDW and IC-CDW−NC-CDW transitions happen at 350 K and 340 K during the heating and cooling process, respectivly. The resistance is measured at low voltage (V=20mV). Optical image of a typical $h$-BN capped 1$T$-TaS$_2$ device is shown in the inset. The dark purple strip is the 1$T$-TaS$_2$ thin film and green larger region is $h$-BN film. The scale bar is 5μm. **c**. Superlinear I-V characteristics below the TS threshold $V_H$. The curve at T=170K is shown here as an example. Arrows indicate the voltage sweeping directions. The inset figure shows that the current below $V_H$ can be fit with an exponential component superimposed on a linear curve. **d**. Temperature dependent TS threshold voltage. Both the $V_H$ and $V_L$ are proportional to (1-T/T$_{NC-IC}$)$^{1/2}$, a dependency identical to that described by analytical theories of the CDW order parameter [1].

G. Liu, B. Debnath, T.R. Pope, T.T. Salguero, R.K. Lake and A. A. Balandin – 2015/2016

**Figure 2**

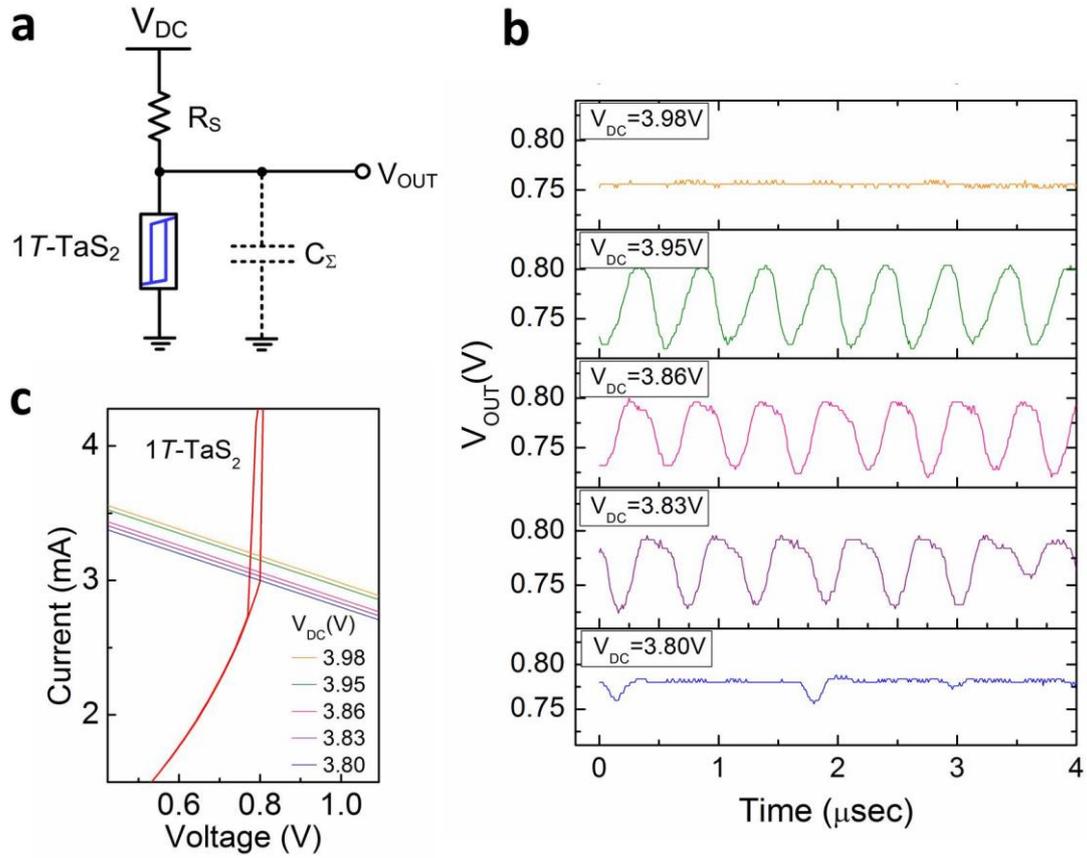

**Figure 2: Oscillator Circuit of 1$T$-TaS$_2$ film with off-chip load resistor**. **a,** Circuit schematic of the oscillator consists of the 1$T$-TaS$_2$ film, a series connected load resistor, and a lumped capacitance from the output node to ground. The load resistance is 1 kΩ. The output terminal is monitored by an oscilloscope. **b**, Voltage oscillations under different V$_{DC}$. The circuit oscillates when V$_{DC}$ is within the range of 3.83-3.95 V. The oscillation frequency is approximately 1.77 MHz, 1.85 MHz, and 2 MHz when V$_{DC}$ is 3.83, 3.86 and 3.95 V, respectively. **c**, Load lines of the resistor at different V$_{DC}$. The blue line, which represents V$_{DC}$=3.8 V, intersects with V$_H$ of 1$T$-TaS$_2$. This is the condition at which the circuit is about to oscillate.

**Figure 3**

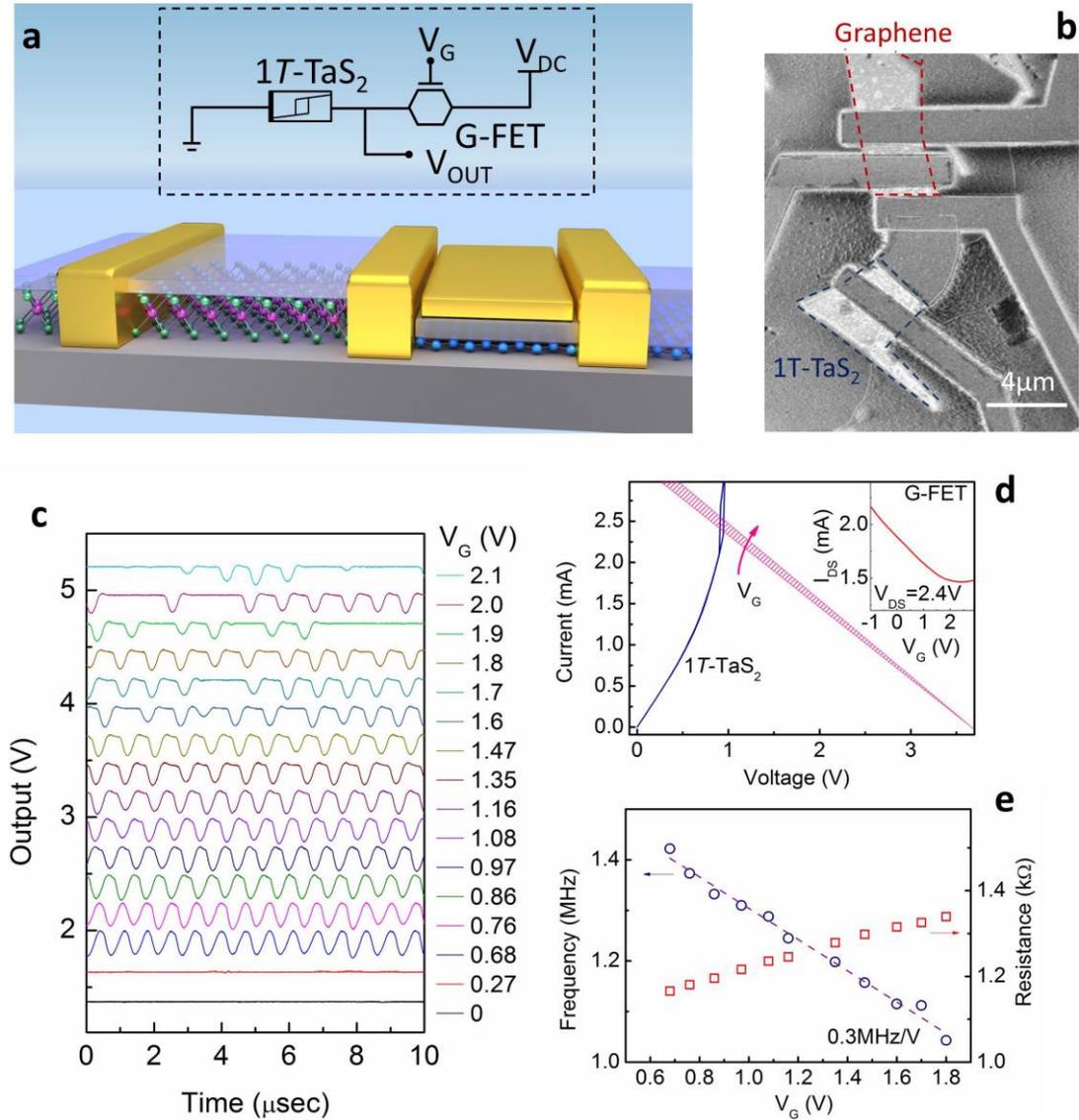

**Figure 3: Integrated 1$T$-TaS$_2$–BN–graphene voltage controlled oscillator. a**, The device structure of the integrated oscillator consists of a graphene FET series connected to 1$T$-TaS$_2$. Both the graphene and TaS$_2$ are fully covered with $h$-BN which acts as a protection layer against oxidation for the 1$T$-TaS$_2$ and as a gate dielectric for the G-FET. The equivalent circuit and biasing condition is shown in the inset. The $V_{DC}$ bias is applied at the drain terminal of the G-FET and the $V_G$ bias is connected to the gate terminal of G-FET. Ground is connected to one terminal of 1$T$-TaS$_2$ device, while the common terminal of the two devices is the output port. **b**, The SEM image of the integrated 1$T$-TaS$_2$–BN–graphene voltage controlled oscillator. The graphene and the TaS$_2$ are highlighted by dashed lines. **c**, Output waveforms at different gate biases when $V_{DC}$ is fixed at 3.65 V. The oscillation frequency is tunable with gate biases in the range of 0.68 V to 1.8 V. The different waveforms are vertically offset of 0.25 V for clarity. **d**, The load line representing the resistance range of the G-FET intersects with the I-V of 1$T$-TaS$_2$ device. The arrow indicates the slope change of the load line with $V_G$. The inset

shows the transfer characteristic ($I_{DS}$ - $V_G$) of G-FET under source drain bias at 2.4 V. **e**, The dependence of oscillation frequency as function of gate bias. Blue circles show the frequency of the oscillation under increased gate bias. The frequency can be adjusted monotonically with the tuning sensitivity of 0.3MHz/V. The red squares are the resistance value of the G-FET under different gate biases with fixed $V_{DC}$=2.4V.